\documentclass[twocolumn,showpacs,preprintnumbers,prb]{revtex4}
\usepackage{graphicx}

\begin{document}

\title{Magnetism and thermodynamics of spin-$(\frac{1}{2},1)$ decorated Heisenberg chain with spin-$1$ pendants}
\author{Shou-Shu Gong, Wei Li, Yang Zhao, Gang Su$^{{\ast }}$}
\affiliation{College of Physical Sciences, Graduate University of Chinese Academy of
Sciences, P. O. Box 4588, Beijing 100049, People's Republic of China}

\begin{abstract}
The magnetic and thermodynamic properties of a new ferrimagnetic
decorated spin-($\frac{1}{2},1$) Heisenberg chain with spin-$1$
pendant spins are investigated for three cases: (A)
$J_{1},J_{2}$$>$$0$; (B) $J_{1}$$>$$0$, $J_{2}$$<$$0$; and (C)
$J_{1}$$<$$0$, $J_{2}$$>$$0$, where $J_{1}$ and $J_{2}$ are the
exchange couplings between spins in the chain and along the rung,
respectively. The low-lying and magnetic properties are explored
jointly by the real-space renormalization group, spin wave, and
density-matrix renormalization group methods, while the
transfer-matrix renormalization group method is invoked to study the
thermodynamics. It is found that the magnon spectra consist of a
gapless and two gapped branches. Two branches in case (C) have
intersections. The coupling dependence of low-energy gaps are
analyzed. In a magnetic field, a $m$=$\frac{3}{2}$ ($m$ is the
magnetization per unit cell) plateau is observed for case (A), while
two plateaux at $m$=$\frac{1}{2}$ and $\frac{3}{2}$ are observed for
cases (B) and (C). Between the two plateaux in cases (B) and (C),
the sublattice magnetizations for the spins coupled by ferromagnetic
interactions have novel decreasing regions with increasing the
magnetic field. At finite temperature, the zero-field susceptibility
temperature product $\chi T$ and specific heat exhibit distinct
exotic features with varying the couplings and temperature for
different cases. $\chi T$ is found to converge as
$T$$\rightarrow$$0$, which is different from the divergent behavior
in the spin-($\frac{1}{2},1$) mixed-spin chain without pendants. The
observed thermodynamic behaviors are also discussed with the help of
their low-lying excitations.

\end{abstract}

\maketitle

\section{Introduction}

In recent years, one-dimensional (1D) quantum ferrimagnets with two
kinds of antiferromagnetically exchange-coupled centers have
attracted much attention due to their exotic properties. The large
families of compounds
ACu(pba)(H$_{2}$O)$_{3}$$\cdot$\textit{n}H$_{2}$O and
ACu(pbaOH)(H$_{2}$O)$_{3}$$\cdot$\textit{n}H$_{2}$O, where A=Mn, Fe,
Co, Ni, Zn, pba=1,3-propylenebis, and
pbaOH=2-hydroxy-1,3-propylenebis, have been extensively explored in
chemistry,\cite{NiCu} which are good realizations of the mixed-spin
chains. These compounds exhibit typically the 1D ferrimagnetic
behavior of $\chi T$ ($\chi$ is the magnetic susceptibility and $T$
is the temperature) that shows a rounded minimum with
temperature.\cite{NiCu}

As a simple model to describe the mixed-spin chains, the
antiferromagnetically coupled spin-($\frac{1}{2}$,$1$) Heisenberg
chain have also been extensively studied by various methods, such as
the spin wave theory,\cite{DMRG1,SW} Schwinger boson mean
field,\cite{SB} density-matrix renormalization group
(DMRG),\cite{DMRG1} quantum Monte Carlo,\cite{QMC} and so
on.\cite{MPS,CP,SO} It has been found that its ground state has a
spontaneous magnetization at $m$=$\frac{1}{2}$ ($m$ is the
magnetization per unit cell) that is consistent with the Lieb-Mattis
theorem,\cite{LM} and the system has a ferrimagnetic (FI) long-range
order. The one-magnon excitation spectra consist of a gapless
ferromagnetic (FM) branch from $S_{G}$ to $S_{G}-1$ ($S_{G}$ is the
good quantum number of total spin in $z$ component in the ground
state) and a gapped antiferromagnetic (AFM) branch from $S_{G}$ to
$S_{G}+1$. \cite{SSH} This magnon gap was numerically found to be
$1.759J$ ($J$ is the exchange coupling).\cite{DMRG1} In a magnetic
field, the system exhibits a magnetization plateau at
$m$=$\frac{1}{2}$ with the width of $1.759J$, corresponding to the
gap of the AFM magnon branch.\cite{ST} Different from the $S$=$1$
Haldane chain, in this mixed-spin chain the spin gap
($1.2795J$) from the ground state to the lowest state in the
subspace of $S_{G}+1$ is less than the magnon gap ($1.759J$) and
thus is not a magnon-like excitation.\cite{DMRG1} The thermodynamic
properties in the coexistence of the AFM and FM
excitations\cite{QMC,TH} and in the critical phase under a magnetic
field\cite{CP,CP2} have also been investigated.

\begin{figure}[tbp]
\includegraphics[width=0.8\linewidth,clip]{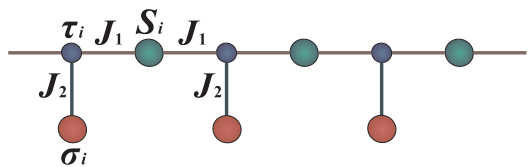}
\caption{(Color online) Sketch of the spin-($\tau,S$) decorated Heisenberg chain with
spin-$\sigma$ pendant spins.}
\label{spin configuration}
\end{figure}

Recently, another interesting family of cyanide-bridged coordination
compounds with pendant magnetic ions are synthesized in
experiment.\cite{CB,ACB} One of them is the cyanide-bridged
Ni(\uppercase\expandafter{\romannumeral2})-Fe(\uppercase\expandafter{\romannumeral3})
complex with a novel building block
[Fe(1-CH$_{3}$im)(CN)$_{5}$]$^{2-}$,\cite{Kou} which can be treated
as the 1D structure as shown schematically in Fig. \ref{spin
configuration} owing to the weak interchain interactions, where the
Ni(\uppercase\expandafter{\romannumeral2}) ($S_{i}$ and
$\sigma_{i}$) and Fe(\uppercase\expandafter{\romannumeral3})
($\tau_{i}$) ions have spin $1$ and $\frac{1}{2}$, respectively.
This compound realizes a decorated spin-($\frac{1}{2},1$) mixed-spin
chain with spin-$1$ pendant spins. Although the intrachain couplings
$J_{1}$$<$$0$ and $J_{2}$$<$$0$ are both FM interactions in the
present compound, it is noticed that any other couplings (i.e.,
$J_{1},J_{2}$$>$$0$, $J_{1}$$>$$0$ and $J_{2}$$<$$0$, $J_{1}$$<$$0$
and $J_{2}$$>$$0$) would give rise to ferrimagnets, making the
realization of such a FI structure more accessible to the
experiment. This family of mixed-spin chains with pendant spins
provides a new scheme to study the 1D quantum ferrimagnetism, which
may have new exotic properties. Although the influences of pendant
spins on magnetism have been discussed in some
antiferromagnets,\cite{AFM} the studies on such ferrimagnets are
still rare till now.

In this paper, we shall explore the physical properties of this new
ferrimagnetic structure, and compare with the spin-($\frac{1}{2},1$)
mixed-spin chain without pendants. The low-lying, magnetic and
thermodynamic properties of the spin-($\frac{1}{2},1$) decorated
Heisenberg chain with spin-$1$ pendant spins for three cases: (A)
$J_{1},J_{2}$$>$$0$; (B) $J_{1}$$>$$0$, $J_{2}$$<$$0$; and (C)
$J_{1}$$<$$0$, $J_{2}$$>$$0$ will be studied using various
techniques. It is unveiled that due to the pendant spins, the new
ferrimagnets exhibit rather distinct magnetic and thermodynamic
behaviors from those of the spin-($\frac{1}{2},1$) mixed-spin chain.
The three cases with different couplings are uncovered to have their
own features, which are expected to observe in experiments. The
exotic properties of this system will also shed light on further
understandings of quantum ferrimagnetism.

This paper is organized as follows. In Sec. \uppercase
\expandafter{\romannumeral2}, the model Hamiltonian is introduced.
In Sec. \uppercase\expandafter{\romannumeral3}, the low-energy
effective Hamiltonians in both strong and
weak couplings are analyzed utilizing the real-space renormalization group
(RSRG) method. The low-lying excitations and magnetic properties are
investigated by the linear spin wave (LSW) theory and DMRG in
Sec. \uppercase\expandafter{\romannumeral4}. In
Sec. \uppercase\expandafter{\romannumeral5}, we shall study the
zero-field thermodynamics by means of the transfer-matrix
renormalization group (TMRG). Finally, a summary and discussion will
be given in Sec. \uppercase\expandafter{\romannumeral6}.

\section{Model Hamiltonian}

The Hamiltonian of the spin-($\frac{1}{2}$,$1$) decorated Heisenberg
chain with spin-$1$ pendant spins in a magnetic field can be written as
\begin{eqnarray}
H&=&\sum^{N}_{i=1}(J_{1}\vec{\tau}_{i}\cdot \vec{S}_{i}
+J_{1}\vec{S}_{i}\cdot \vec{\tau}_{i+1}
+J_{2}\vec{\tau}_{i}\cdot \vec{\sigma}_{i}) \nonumber \\
&-&h\sum^{N}_{i=1}(\tau^{z}_{i}+S^{z}_{i}+\sigma^{z}_{i}),
\label{model Hamiltonian}
\end{eqnarray}
where $\vec{\sigma}_{i}$ is the $\sigma$=$1$ pendant spin,
$\vec{\tau}_{i}$ and $\vec{S}_{i}$ are the spins in the chain with
$\tau$=$\frac{1}{2}$ and $S$=$1$, respectively, $J_{1,2}$$>$$0$
($<$$0$) denote the AFM (FM) couplings, and $h$ is the magnetic
field. Throughout the context, we take $J_{1}$ as an energy scale
and $g\mu _{B}$=$1$. The schematic representation of the model is
shown in Fig. \ref{spin configuration}.

Analogous to the spin-($\frac{1}{2}$,$1$) mixed-spin chain without pendants, the
system with Hamiltonian (\ref{model Hamiltonian}) has a
spontaneous magnetization in the absence of magnetic field according
to the Lieb-Mattis theorem.\cite{LM} In case (A) ($J_{1,2}$$>$$0$) the spontaneous
magnetization per unit cell is $m$=$\frac{3}{2}$, while in both cases (B)
($J_{1}$$>$$0$, $J_{2}$$<$$0$) and (C) ($J_{1}$$<$$0$,
$J_{2}$$>$$0$) it is spontaneously magnetized at $m$=$\frac{1}{2}$.
The Goldstone theorem\cite{GT} allows gapless
excitations in these cases owing to the spontaneous breaking of the SU(2)
symmetry.

\section{Real Space renormalization group analysis}

In this section, the low-energy effective Hamiltonians of the three
cases in both strong- and weak-coupling limits are derived utilizing
the RSRG.\cite{RSRG} In the RSRG procedure, the Hamiltonian is
divided into intrablock ($H^{B}$) and interblock ($H^{BB}$) parts.
By diagonalizing $H^{B}$, a number of low-energy states are kept
to project the full Hamiltonian into the renormalized Hilbert space.
Although RSRG cannot give the results as accurate as the numerical
approaches, it can give a good qualitative description for low-energy
properties.

\subsection{$J_{1}$$>$$0$, $J_{2}$$>$$0$}

Let us first consider the strong-coupling limit
($J_{2}$$\gg$$J_{1}$). Since the interaction between $\tau_{i}$ and
$\sigma_{i}$ is strong, each rung can be considered as the isolated
block in the first step of renormalization group (RG). Each block
consists of two multiplets whose total spins are $1/2$ and $3/2$
with energies $-J_{2}$ and $J_{2}/2$, respectively. The
spin-$\frac{1}{2}$ doublets are kept as the basis to construct the
embedding operator $T$ to project the full Hamiltonian onto the
truncated Hilbert space. The effective Hamiltonian can be obtained
as
\begin{equation}
\tilde{H}_{\mathrm{A}}^{\mathrm{strong},1}=-NJ_{2}-\frac{1}{3}J_{1}\sum^{N}_{i=1}(\vec{S}^{\prime}_{i}
\cdot \vec{S}_{i}+\vec{S}_{i}\cdot \vec{S}^{\prime}_{i+1}),
\label{AAS1}
\end{equation}
where $S^{\prime}_{i}$=$1/2$ is the renormalized spin truncated from
the rung block. The Hamiltonian (\ref{AAS1}) describes a
spin-($\frac{1}{2},1$) mixed-spin chain with a renormalized FM
coupling $-\frac{1}{3}J_{1}$. In the next step of RG procedure, the
Hamiltonian (\ref{AAS1}) is further projected onto a
$S^{\prime\prime}$=$3/2$ FM Heisenberg chain
\begin{equation}
\tilde{H}_{\mathrm{A}}^{\mathrm{strong},2}=-(J_{2}+\frac{J_{1}}{6})N-\frac{2}{27}J_{1}\sum_{i=1}^{N}\vec{S}^{\prime\prime}_{i}\cdot \vec{S}^{\prime\prime}_{i+1}.
\label{AAS2}
\end{equation}
The magnon excitations in this FM Hamiltonian correspond to the
magnons from $S_{G}$ to $S_{G}-1$ in the original system, which are
hence expected to be gapless with a quadratic dispersion in
low energies. The RG can also give the sublattice magnetization
$m_{\mathrm{S}}$=$1$, $m_{\mathrm{\tau}}$=$-\frac{1}{6}$, and
$m_{\mathrm{\sigma}}$=$\frac{2}{3}$. The sum of them gives
$\frac{3}{2}$, recovering the spontaneous magnetization of the
original system.

In the weak-coupling limit ($J_{2}$$\ll$$J_{1}$), the spins
$\vec{\tau}_{i}$ and $\vec{S}_{i}$ are taken as a block, and the
doublets with spin $1/2$ are kept to truncate the block Hilbert
space. The effective Hamiltonian is
\begin{equation}
\tilde{H}_{\mathrm{A}}^{\mathrm{weak},1}=-NJ_{1}-\sum_{i=1}^{N}(\frac{4}{9}J_{1}\vec{S}^{\prime}_{i}\cdot \vec{S}^{\prime}_{i+1}+\frac{1}{3}J_{2}\vec{S}^{\prime}_{i}\cdot \vec{\sigma}_{i}),
\label{AAW1}
\end{equation}
where $S^{\prime}_{i}$=$1/2$ is the renormalized block spin. The
Hamiltonian is mapped onto a spin-$1/2$ FM Heisenberg chain with
$\sigma$=$1$ pendants coupled by the renormalized FM
interaction $-\frac{1}{3}J_{2}$. The spin wave analysis unveils that
it has a gapless FM excitation with the dispersion
\begin{equation}
\omega_{k}\sim\frac{2}{27}J_{1}k^{2}
\label{ZeroD1}
\end{equation}
for $k$$\rightarrow$$0$, which corresponds to the magnon excitation
from $S_{G}$ to $S_{G}-1$ of the original system. The sublattice
magnetization is obtained as $m_{\mathrm{S}}$=$\frac{2}{3}$,
$m_{\mathrm{\tau}}$=$-\frac{1}{6}$, and $m_{\mathrm{\sigma}}$=$1$,
and the sum of them is also $\frac{3}{2}$.

From Eqs. (\ref{AAS1}) and (\ref{ZeroD1}) it can be seen that in the
two coupling limits, the low-energy behaviors of the gapless branch
are both dominated by $J_{1}$. Besides, $m_{\mathrm{S}}$ and
$m_{\mathrm{\sigma}}$ exchange their values in the two limits, and
$m_{\tau}$ is unchanged, implying a possible crossing of $m_{\tau}$
in the intermediate region of $J_{2}/J_{1}$, which would be
confirmed by the DMRG results in the next section.

\subsection{$J_{1}$$>$$0$, $J_{2}$$<$$0$}

In the strong-coupling limit ($\vert J_{2}\vert$$\gg$$J_{1}$), owing to the strong FM $J_{2}$, the low-energy
multiplets with total spin $3/2$ are kept to project the Hamiltonian
in the first step of RG. The effective Hamiltonian is obtained as
\begin{equation}
\tilde{H}_{\mathrm{B}}^{\mathrm{strong},1}=\frac{1}{2}J_{2}N+\frac{1}{3}J_{1}\sum_{i=1}^{N}(\vec{S}^{\prime}_{i}\cdot \vec{S}_{i}+\vec{S}_{i}\cdot \vec{S}^{\prime}_{i+1})
\label{AFS1}
\end{equation}
with the renormalized spin of rung block $S^{\prime}_{i}$=$3/2$, which
depicts a spin-($\frac{3}{2},1$)
Heisenberg chain with a renormalized AFM coupling $\frac{1}{3}J_{1}$.
In the next step of RG procedure the Hamiltonian (\ref{AFS1}) is
projected to a $S^{\prime\prime}$=$1/2$ FM Heisenberg chain
\begin{equation}
\tilde{H}_{\mathrm{B}}^{\mathrm{strong},2}=\frac{1}{6}(3J_{2}-5J_{1})N-\frac{10}{27}J_{1}\sum_{i=1}^{N}\vec{S}^{\prime\prime}_{i}\cdot \vec{S}^{\prime\prime}_{i+1},
\label{AFS2}
\end{equation}
whose FM excitations imply that the magnon excitations of the
original system from $S_{G}$ to $S_{G}-1$ are gapless with a
quadratic dispersion relation in low energies. The sublattice
magnetization is obtained as $m_{\mathrm{S}}$=$-\frac{1}{3}$,
$m_{\mathrm{\tau}}$=$\frac{5}{18}$, and
$m_{\mathrm{\sigma}}$=$\frac{5}{9}$, giving the spontaneous
magnetization per unit cell $m$=$\frac{1}{2}$.

In the weak-coupling limit ($\vert J_{2}\vert$$\ll$$J_{1}$), we
perform the RG on the block spins $\vec{S}_{i}$ and $\vec{\tau}_{i}$
with the doublet of total spin $1/2$. The effective Hamiltonian is
\begin{equation}
\tilde{H}_{\mathrm{B}}^{\mathrm{weak},1}=-NJ_{1}-\sum_{i=1}^{N}(\frac{4}{9}J_{1}\vec{S}^{\prime}_{i}\cdot \vec{S}^{\prime}_{i+1}+\frac{1}{3}J_{2}\vec{S}^{\prime}_{i}\cdot \vec{\sigma}_{i})
\label{AFW1}
\end{equation}
with the renormalized spin $S^{\prime}_{i}$=$1/2$, which describes a
spin-$1/2$ FM Heisenberg chain with $\sigma$=$1$ pendant spins
coupled by the renormalized AFM interaction $-\frac{1}{3}J_{2}$. The
spin wave results show that the excitations that correspond to those
from $S_{G}$ to $S_{G}-1$ in the original system are also gapless
with
\begin{equation}
\omega_{k}\sim\frac{2}{9}J_{1}k^{2}
\label{ZeroD2}
\end{equation}
for $k$$\rightarrow$$0$. In the two limits, it is observed from Eqs.
(\ref{AFS1}) and (\ref{ZeroD2}) that the low-energy behaviors of the
gapless excitation are both dominated by $J_{1}$.

\subsection{$J_{1}$$<$$0$, $J_{2}$$>$$0$}

For the strong-coupling limit ($J_{2}$$\gg$$\vert J_{1}\vert$), because of
the strong AFM $J_{2}$, $\vec{\sigma}_{i}$ and
$\vec{\tau}_{i}$ are renormalized by the doublet with total spin
$1/2$. The effective Hamiltonian is given by
\begin{equation}
\tilde{H}_{\mathrm{C}}^{\mathrm{strong},1}=-NJ_{2}-\frac{1}{3}J_{1}\sum^{N}_{i=1}(\vec{S}^{\prime}_{i}
\cdot \vec{S}_{i}+\vec{S}_{i}\cdot \vec{S}^{\prime}_{i+1})
\label{FAS1}
\end{equation}
with $S^{\prime}$=$1/2$. Eq. (\ref{FAS1}) describes a
spin-($\frac{1}{2},1$) Heisenberg chain coupled by the renormalized
AFM interaction $-\frac{1}{3}J_{1}$. In the second step of RG, the
Hamiltonian (\ref{FAS1}) is projected to a $S^{\prime\prime}$=$1/2$
FM Heisenberg chain
\begin{equation}
\tilde{H}_{\mathrm{C}}^{\mathrm{strong},2}=-(J_{2}-\frac{J_{1}}{3})N+\frac{4}{27}J_{1}\sum_{i=1}^{N}\vec{S}^{\prime\prime}_{i}\cdot \vec{S}^{\prime\prime}_{i+1},
\label{FAS2}
\end{equation}
which unveils the gapless excitations from $S_{G}$ to $S_{G}-1$ of
the original system. The sublattice magnetization is obtained as
$m_{\mathrm{S}}$=$\frac{2}{3}$, $m_{\mathrm{\tau}}$=$\frac{1}{18}$,
and $m_{\mathrm{\sigma}}$=$-\frac{2}{9}$, giving rise to the
spontaneous magnetization per unit cell $m$=$\frac{1}{2}$.

In the weak-coupling limit ($J_{2}$$\ll$$\vert J_{1}\vert$), the
spins $\vec{S}_{i}$ and $\vec{\tau}_{i}$ are taken as a block, and
the multiplet with spin $3/2$ are kept to truncate the block Hilbert
space. The effective Hamiltonian is obtained as
\begin{equation}
\tilde{H}_{\mathrm{C}}^{\mathrm{weak},1}=\frac{1}{2}NJ_{1}+\sum_{i=1}^{N}(\frac{2}{9}J_{1}\vec{S}^{\prime}_{i}\cdot \vec{S}^{\prime}_{i+1}+\frac{1}{3}J_{2}\vec{S}^{\prime}_{i}\cdot \vec{\sigma}_{i}),
\label{FAW1}
\end{equation}
which describes a spin-$3/2$ FM Heisenberg chain with
antiferromagnetically coupled pendant spins $\sigma_{i}$. The spin
wave analysis indicates that the effective system has a FM gapless
excitation with
\begin{equation}
\omega_{k}\sim-J_{1}k^{2}
\label{ZeroD3}
\end{equation}
for $k$$\rightarrow$$0$. Analogous to the above cases, the low-energy
behavior of the gapless branch in the two limits are determined by
$J_{1}$, which can be seen from Eqs. (\ref{FAS1}) and (\ref{ZeroD3}).

Based on the RSRG analyses, one may observe that the different cases
have distinct low-energy effective Hamiltonians, and the magnon
excitations from $S_{G}$ to $S_{G}-1$ are always FM and gapless,
being consistent with those of the spin-($\frac{1}{2},1$) mixed spin
chain. The dispersion relations near $k$=$0$ are found to be
dominated by $J_{1}$ in both strong and weak couplings. The
low-energy effective Hamiltonians for cases (B) and (C) in the
strong-coupling limit are analogous except the magnitude of spin,
whose thermodynamic properties will be compared in Sec.
\uppercase\expandafter{\romannumeral5}.

\section{Low-lying excitations and magnetic properties}

In this section, the low-lying excitations and magnetic properties
are explored by means of the LSW\cite{DMRG1,SW} and
DMRG.\cite{DMRG2,DMRG3} During the DMRG calculations, the chain length is
taken as $L$=$300$, and the Hilbert space is truncated to $240$ most
relevant states. Open boundary conditions are adopted, and the
truncation error is less than $10^{-8}$ in all calculations.

\subsection{$J_{1}$$>$$0$, $J_{2}$$>$$0$}

\begin{figure}[tbp]
\includegraphics[width=1.0\linewidth,clip]{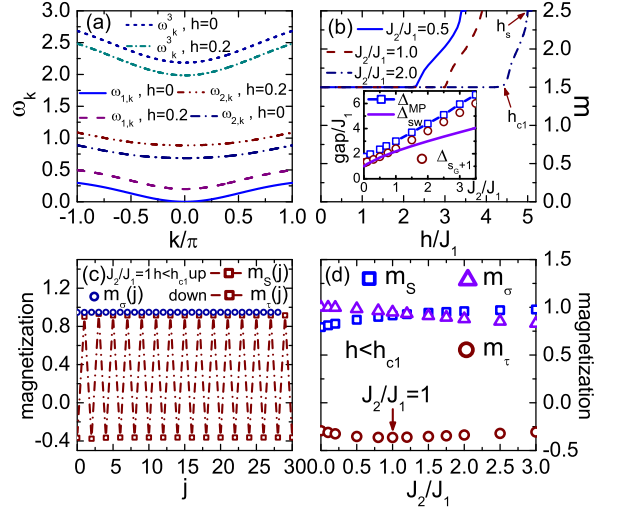}
\caption{(Color online) (a) Magnon excitation dispersion for case (A) with
$J_{1}$=$J_{2}$=$1$ under different magnetic fields. (b) Magnetization curves for different
$J_{2}$. The inset shows the coupling dependence of the gap
$\Delta_{\mathrm{MP}}$, $\Delta_{\mathrm{SW}}$, and
$\Delta_{\mathrm{S_{G}+1}}$. (c) Local magnetization as a function
of lattice site for $h$$<$$h_{c1}$. (d) Coupling dependence of
sublattice magnetization for $J_{2}$/$J_{1}$=$1$ and $h$$<$$h_{c}$.} \label{AF-AF}
\end{figure}

The Holstein-Primakoff (HP) transformations are introduced as follows:
\begin{eqnarray}
\sigma^{z}_{i}&=&s_{1}-a^{\dagger}_{i}a_{i}, \nonumber \\
\sigma^{+}_{i}&=&\sqrt{2s_{1}-a^{\dagger}_{i}a_{i}}a_{i}, \nonumber \\
\sigma^{-}_{i}&=&a^{\dagger}_{i}\sqrt{2s_{1}-a^{\dagger}_{i}a_{i}}
\label{HP}
\end{eqnarray}
for the sublattice of $\vec{\sigma}_{i}$ spins with $s_{1}$=$1$, and
\begin{eqnarray}
\tau^{z}_{i}&=&-s_{2}+b^{\dagger}_{i}b_{i}, \nonumber \\
\tau^{+}_{i}&=&b^{\dagger}_{i}\sqrt{2s_{2}-b^{\dagger}_{i}b_{i}}, \nonumber \\
\tau^{-}_{i}&=&\sqrt{2s_{2}-b^{\dagger}_{i}b_{i}}b_{i} \label{HP2}
\end{eqnarray}
for the sublattice of spins $\vec{\tau}_{i}$ with
$s_{2}$=$\frac{1}{2}$, where the operators $a_{i}$ and $b_{i}$ are
bosons. The spins $\vec{S}_{i}$ are transformed in the similar way
as Eq. (\ref{HP}) with bosonic operators $c_{i}$ and
$c^{\dagger}_{i}$. Thus, the magnon spectra can be obtained by
diagonalizing the Hamiltonian after performing the Fourier and
Bogoliubov transformations. As shown in Fig. \ref{AF-AF}(a), the
magnon spectra consist of a gapless branch $\omega_{1,k}$ and two
gapped ones $\omega_{2,k}$ and $\omega_{3,k}$. In the presence of a
magnetic field $h$, both $\omega_{1,k}$ and $\omega_{2,k}$ increase,
while $\omega_{3,k}$ decreases, indicating that $\omega_{1,k}$ and
$\omega_{2,k}$ describe the magnons from $S_{G}$ to $S_{G}-1$ while
$\omega_{3,k}$ are those from $S_{G}$ to $S_{G}+1$. For $J_{2}$=$0$,
the spectra are reduced to a gapped and a gapless excitations, which
agree exactly with those of the spin-($\frac{1}{2}$,$1$) mixed-spin
chain. When $J_{2}$ is set in, the gapless branch splits into
$\omega_{1,k}$ and $\omega_{2,k}$. With increasing $J_{2}$, both
$\omega_{2,k}$ and $\omega_{3,k}$ enhance. It is found that the
low-energy dispersions near $k$=$0$ of the gapless branch
$\omega_{1,k}$ are insensitive to $J_{2}$ but dominated by $J_{1}$
in a wide range of the coupling ratio, which covers the result
obtained from the RSRG.

The magnetic curve $m(h)$ and low-energy gaps are then studied by
the DMRG method. As shown in Fig. \ref{AF-AF}(b), $m(h)$ has a
plateau at the spontaneous magnetization $m$=$\frac{3}{2}$, whose
width $\Delta_{\mathrm{MP}}$ increases with increasing $J_{2}$. In
this figure, $h_{c1}$ denotes the field where the plateau
disappears, and $h_{s}$ is the saturation field. For $J_{2}$=$0$,
$\Delta_{\mathrm{MP}}$ reduces to $1.759J_{1}$ of the
spin-($\frac{1}{2}$,$1$) mixed-spin chain, which is exactly the gap
of its massive magnon branch.\cite{SSH,ST} The coupling dependence
of $\Delta_{\mathrm{MP}}$ is illustrated in the inset of Fig.
\ref{AF-AF}(b), showing that $\Delta_{\mathrm{MP}}$ increases almost
as a linear behavior. The $J_{1}$ dependence of
$\Delta_{\mathrm{MP}}$ is also studied by taking $J_{2}$ as the
energy scale, which is not presented here. It is found that
$\Delta_{\mathrm{MP}}$/$J_{2}$ varies rather slowly with $J_{1}$,
which means that $\Delta_{\mathrm{MP}}$ is mainly scaled by $J_{2}$
in this case. The gap of the massive magnon branch $\omega_{3,k}$
($\Delta_{\mathrm{SW}}$) is also shown in the inset of Fig.
\ref{AF-AF}(b) in comparison to $\Delta_{\mathrm{MP}}$. The magnon
gap obtained from the LSW appears to be smaller than
$\Delta_{\mathrm{MP}}$, where the deviation increases for stronger
$J_{2}$. It appears that the LSW underestimates the magnon gap from
$S_{G}$ to $S_{G}+1$.

We also compute the spin gap $\Delta_{\mathrm{S_{G}+1}}$ from the
ground state to the lowest state in the $S_{G}+1$ subspace, as shown
in the inset of Fig. \ref{AF-AF}(b). Analogous to the
spin-($\frac{1}{2}$,$1$) mixed-spin chain,
$\Delta_{\mathrm{S_{G}+1}}$ is smaller than $\Delta_{\mathrm{MP}}$,
indicating that $\Delta_{\mathrm{S_{G}+1}}$ is also not a
magnon-like excitation. But, it has a similar behavior with coupling
to $\Delta_{\mathrm{MP}}$, which can thus be used to describe the
low-energy behaviors. The spin gap from the ground state to the
lowest state in the $S_{G}-1$ subspace is computed, which is found
always vanishing and is consistent with the gapless branch
$\omega_{1,k}$.

Next let us discuss the spin-spin correlation function and local
magnetization in ground states. Figure \ref{AF-AF}(c) shows the
local magnetization as a function of lattice site for
$h$$<$$h_{c1}$. In the ground state, the spin correlation functions
along the chain have a long-range order (LRO), and the spin
fluctuations $\langle S^{z}_{i}S^{z}_{j}\rangle$-$\langle
S^{z}_{i}\rangle$$\langle S^{z}_{j}\rangle$ decay rather rapidly
(not presented here). Hence it is adequate to study only the local
magnetization in the ground state. For $J_{2}$=$0$, it gives
$m_{\mathrm{\tau}}$=$-0.29248$ and $m_{\mathrm{S}}$=$0.79248$,
exactly in agreement with the previous result. \cite{DMRG1} After
tuning on $J_{2}$, the coupling dependence of sublattice
magnetization is shown in Fig. \ref{AF-AF}(d). The pendant spin
magnetization $m_{\mathrm{\sigma}}$ is suppressed by the quantum
fluctuations after tuning $J_{2}$, while $m_{\mathrm{S}}$ increases
and approaches saturation for large $J_{2}$, confirming the results
of the RSRG. The interesting phenomenon is the behavior of
$m_{\mathrm{\tau}}$. As shown by the arrow in Fig. \ref{AF-AF}(d),
$m_{\mathrm{\tau}}$ decreases for $J_{2}$/$J_{1}$$<$$1$, and turns
to increase when $J_{2}$/$J_{1}$$>$$1$, which has a turning point at
$J_{2}$/$J_{1}$$=$$1$, as suggested by the result of the RSRG.
Meanwhile, it can be seen that $m_{\mathrm{\sigma}}$ and
$m_{\mathrm{S}}$ intersect near $J_{2}$/$J_{1}$$=$$1$. The changes
of coupling dependence of the magnetic moments near
$J_{2}$/$J_{1}$$=$$1$ may be owing to the competition of the two AFM
interactions.

\subsection{$J_{1}$$>$$0$, $J_{2}$$<$$0$}

\begin{figure}[tbp]
\includegraphics[width=1.0\linewidth,clip]{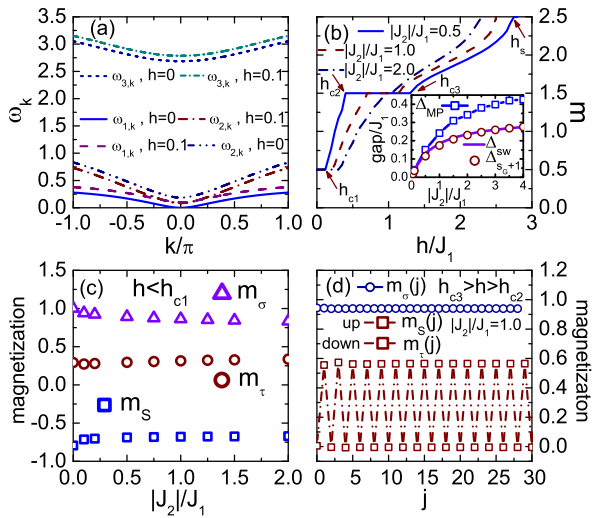}
\caption{(Color online) (a) Magnon excitation dispersion for case (B) with
$J_{1}$=$1$ and $J_{2}$=$-1$. (b) Magnetization curves for different
$J_{2}$. The inset shows the coupling dependence of the gap
$\Delta_{\mathrm{MP}}$, $\Delta_{\mathrm{SW}}$, and
$\Delta_{\mathrm{S_{G}+1}}$. (c) Coupling dependence of sublattice
magnetization. (d) Local magnetization as a function of lattice
site for $h_{c2}$$<$$h$$<$$h_{c3}$.} \label{AF-F}
\end{figure}

To perform the LSW calculation, $\vec{\mathbf{\sigma}}_{i}$ and
$\vec{\tau}_{i}$ spins are transformed as the form of Eq. (\ref{HP})
by the bosonic operators ($a_{i}$,$a^{\dagger}_{i}$) and
($b_{i}$,$b^{\dagger}_{i}$) with $s_{1}$=$1$ and $\frac{1}{2}$,
respectively, while $\vec{S}_{i}$ are transformed as the form of Eq.
(\ref{HP2}) by ($c_{i}$,$c^{\dagger}_{i}$) with $s_{2}$=$1$. As
shown in Fig. \ref{AF-F}(a), the spectra consist of a gapless and
two gapped branches. In the presence of magnetic field, the gapless
branch $\omega_{1,k}$ and the gapped branch $\omega_{3,k}$ increase,
while the gapped branch $\omega_{2,k}$ decreases, indicating that
$\omega_{1,k}$ and $\omega_{3,k}$ are the excitations from the
sector $S_{G}$ to $S_{G}-1$ while $\omega_{2,k}$ describes the
excitations from $S_{G}$ to $S_{G}+1$. It can be seen that the
spectra in this case are quite different from those of case (A). The
branch $\omega_{2,k}$ from $S_{G}$ to $S_{G}+1$ is close to the
gapless branch $\omega_{1,k}$ for the present case, and with
increasing $|J_{2}|$, it increases slightly. The resulting
distinctions in the thermodynamics would be explored in Sec.
\uppercase\expandafter{\romannumeral5}. With changing the couplings,
it is found that the low-energy dispersions of the gapless branch
$\omega_{1,k} $ near $k$=$0$ are dominated by $J_{1}$, which is
consistent with the RSRG result.

The magnetic curve $m(h)$ and low-energy gaps are shown in Fig.
\ref{AF-F}(b). It can be seen that $m(h)$ exhibits two plateaux at
$m$=$\frac{1}{2}$ and $\frac{3}{2}$. We denote the field where the
$m$=$\frac{1}{2}$ plateau vanishes as $h_{c1}$, the lower and upper
critical fields for the $m$=$\frac{3}{2}$ plateau as $h_{c2}$ and
$h_{c3}$, respectively, and $h_{s}$ as the saturation field. With
increasing $\vert J_{2}\vert$, the width of the $m$=$\frac{1}{2}$
plateau $\Delta_{\mathrm{MP}}$ is enlarged slightly, while the
$m$=$\frac{3}{2}$ plateau decreases and smears when $\vert
J_{2}\vert$/$J_{1}$$\gtrsim$$2.0$. It should be noted that in some
quasi-one dimensional polymerized Heisenberg antiferromagnets, there
might be a transition from the plateau state to the non-plateau
state that is usually of the Kosterlitz-Thouless type.\cite{KT} Such
a transition point cannot be numerically determined accurately owing
to the finite-size length of the chain. Therefore, in the present
case, whether or not a plateau-non-plateau transition with couplings
at the $m$=$\frac{3}{2}$ plateau exists is still uncertain from our
DMRG numerical results. The coupling dependence of
$\Delta_{\mathrm{MP}}$ is illustrated in the inset of Fig.
\ref{AF-F}(b), showing that $\Delta_{\mathrm{MP}}$ increases with
enhancing $\vert J_{2}\vert$, and different from case (A),
$\Delta_{\mathrm{MP}}$ goes to saturate at large $\vert J_{2}\vert$,
which suggests that $\Delta_{\mathrm{MP}}$ is mainly scaled by
$J_{1}$ for large $\vert J_{2}\vert$/$J_{1}$. The coupling
dependence of the magnon gap $\omega_{2,k=0}$
($\Delta_{\mathrm{SW}}$) is also shown in the inset of Fig.
\ref{AF-F}(b). It can be seen that the spin wave is capable of
describing the coupling dependence of magnon gap qualitatively,
though it underestimates the value like in the case (A). The spin
gap $\Delta_{\mathrm{S_{G}+1}}$ from the ground state to the lowest
state in the $S_{G}+1$ subspace is also computed. As shown in the
inset of Fig. \ref{AF-F}, $\Delta_{\mathrm{S_{G}+1}}$ is less than
$\Delta_{\mathrm{MP}}$, indicating that it is also not a magnon-like
excitation, and its behavior for different coupling ratios is
consistent with $\Delta_{\mathrm{MP}}$ and $\Delta_{\mathrm{SW}}$.

The coupling dependence of sublattice magnetization in the ground
states is shown in Fig. \ref{AF-F}(c). As the quantum fluctuations
become strong after tuning $J_{2}$, the pendant spin magnetization
$m_{\mathrm{\sigma}}$ decreases with increasing $\vert J_{2}\vert$,
which is analogous to case (A). However, the FM coupling has
different effects on the spins in the chain compared with the AFM
$J_{2}$. With increasing $\vert J_{2}\vert$, both
$m_{\mathrm{\tau}}$ and $m_{\mathrm{S}}$ increase slightly, and
$m_{\mathrm{\tau}}$ does not show an extremum like in the case (A).

\begin{figure}[tbp]
\includegraphics[width=1.0\linewidth,clip]{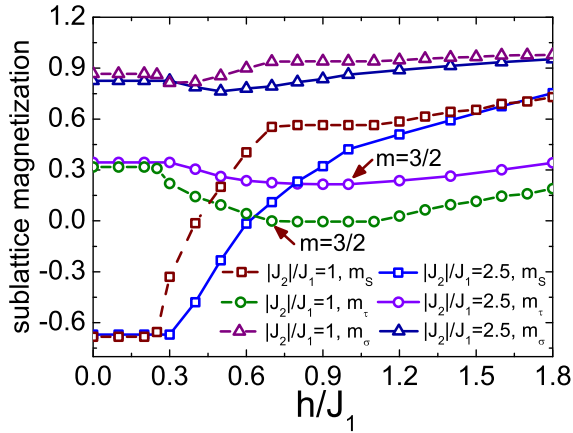}
\caption{(Color online) Magnetic field dependence of the sublattice
magnetization $m_{\mathrm{S}}$, $m_{\mathrm{\tau}}$ and
$m_{\mathrm{\sigma}}$ for $\vert J_{2}\vert$/$J_{1}$=$1$ and $2.5$.
The arrows indicate the minimum of $m_{\mathrm{\tau}}$, where the
magnetization per unit cell is $m$=$3/2$.} \label{Mag}
\end{figure}

In the $m$=$\frac{3}{2}$ plateau region ($h_{c2}$$<$$h$$<$$h_{c3}$),
the local magnetization is shown in Fig. \ref{AF-F}(d) for $\vert
J_{2}\vert/J_{1}$=$1$ as an example. It may be expected that all the
local magnetic moments would increase from $h_{c1}$ to $h_{c2}$.
However, by comparing the local magnetization below $h_{c1}$ [Fig.
\ref{AF-F}(c)] and in the plateau [Fig. \ref{AF-F}(d)], it is
surprising to notice that $m_{\mathrm{\tau}}$ decreases from
$0.3156$ below $h_{c1}$ to $-0.0046$ in $h_{c2}$. Therefore, the
field dependence of sublattice magnetization is studied, as shown by
$m_{\mathrm{S}}$, $m_{\mathrm{\tau}}$, and $m_{\mathrm{\sigma}}$ for
$\vert J_{2}\vert$/$J_{1}$=$1$ and $2.5$ in Fig. \ref{Mag}. It can
be seen that $m_{\mathrm{\tau}}$ decreases continuously from
$h_{c1}$ to $h_{c2}$, while $m_{\mathrm{\sigma}}$ decreases in a
short range above $h_{c1}$. For a comparison, we also calculated the
sublattice magnetization as a function of field for both the
spin-($\frac{1}{2},1$) mixed-spin chain and the case (A). The
results shows that the above behavior is not seen. Therefore, this
novel decreasing behavior may be owing to the competition between
the FM and AFM interactions in a magnetic field.

\subsection{$J_{1}$$<$$0$, $J_{2}$$>$$0$}

\begin{figure}[tbp]
\includegraphics[width=1.0\linewidth,clip]{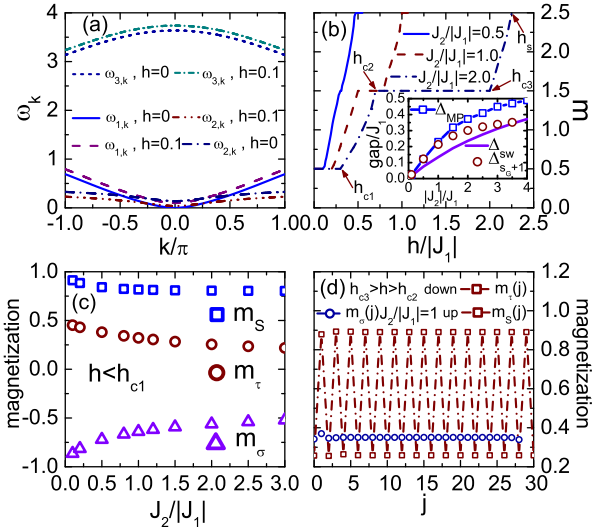}
\caption{(Color online) (a) Magnon excitation dispersion for case (C) with
$J_{1}$=$-1$ and $J_{2}$=$1$. (b) Magnetization curves for different
$J_{2}$. The inset shows the coupling dependence of the gap
$\Delta_{\mathrm{MP}}$, $\Delta_{\mathrm{SW}}$, and
$\Delta_{\mathrm{S_{G}+1}}$. (c) Coupling dependence of sublattice
magnetization. (d) Local magnetization as a function of lattice
site for $h_{c2}$$<$$h$$<$$h_{c3}$.} \label{F-AF}
\end{figure}

For $J_{1}$$<$$0$ and $J_{2}$$>$$0$, the HP transformations are
applied on the spins $\vec{S}_{i}$ and $\vec{\tau}_{i}$ with the
form of Eq. (\ref{HP}) by ($b_{i}$,$b^{\dagger}_{i}$) and
($c_{i}$,$c^{\dagger}_{i}$) for $s_{1}$=$1$ and $\frac{1}{2}$,
respectively, and that with the form of Eq. (\ref{HP2}) is applied
on $\vec{\sigma}_{i}$ for $s_{2}$=$1$ with
($a_{i}$,$a^{\dagger}_{i}$). As shown in Fig. \ref{F-AF}(a), the
spectra consist of a gapless ($\omega_{1,k}$) and a gapped
($\omega_{3,k}$) magnon branches from the sector $S_{G}$ to
$S_{G}-1$, as well as a gapped one ($\omega_{2,k}$) from $S_{G}$ to
$S_{G}+1$, which can be identified by the shifts of the branches
with the magnetic field. It is noticed that the gapped branch
$\omega_{2,k}$ from $S_{G}$ to $S_{G}+1$ is close to the gapless
branch $\omega_{1,k}$, which is similar to the case (B), but
$\omega_{2,k}$ has lower energies than $\omega_{1,k}$ for large wave
momenta $k$ in the present case. With increasing $J_{2}$/$\vert
J_{1}\vert$, $\omega_{2,k}$ enhances and the intersected momenta of
the two branches shift to higher values. A similar intersection of
magnon branches has also been observed in the
spin-($\frac{1}{2}$,$1$) mixed-spin chain with AFM nearest-neighbor
and FM next-nearest-neighbor interactions.\cite{SSKS} The influences
of this intersection on the thermodynamics would be discussed in the
next section. For $\omega_{1,k}$, it is found that the low-energy
dispersions near $k$=$0$ are also dominated by $J_{1}$, which agrees
with the RSRG analysis.

In Fig. \ref{F-AF}(b), the magnetic curves $m(h)$ for different
couplings are shown. Similar to case (B), $m(h)$ has two plateaux at
$m$=$\frac{1}{2}$ and $\frac{3}{2}$, whose critical fields are
denoted by the same symbols as the case (B). With increasing
$J_{2}$/$\vert J_{1}\vert$, the width of the $m$=$\frac{1}{2}$
plateau ($\Delta_{\mathrm{MP}}$) extends, while that of the
$m$=$\frac{3}{2}$ plateau is enlarged, which differs from the case
(B) where the $m$=$\frac{3}{2}$ plateau decreases with increasing
the coupling ratio. The inset of Fig. \ref{F-AF}(b) shows the
coupling dependence of the low-energy gaps $\Delta_{\mathrm{MP}}$,
$\Delta_{\mathrm{SW}}$, and $\Delta_{\mathrm{S_{G}+1}}$. It can be
seen that the gaps behave similarly to those in the case (B). The
gaps approach to the saturation for large $J_{2}$/$\vert
J_{1}\vert$, indicating that they are mainly scaled by $J_{1}$ in
the large $J_{2}$ limit. The LSW also underestimates the magnon gap
of $\omega_{2,k}$ as $\Delta_{\mathrm{SW}}$ is smaller than
$\Delta_{\mathrm{MP}}$. $\Delta_{\mathrm{S_{G}+1}}$ appears to be
smaller than $\Delta_{\mathrm{MP}}$, which means that the spin gap
from the ground state to the lowest state in the subspace with
$S_{G}+1$ is also not a magnon-like excitation. In the ground
states, the coupling dependence of the sublattice magnetization is
displayed in Fig. \ref{F-AF}(c). It can be seen that as the quantum
fluctuations are induced by $J_{2}$, $m_{\mathrm{S}}$ and $m_{\tau}$
decrease with increasing $J_{2}$, while $m_{\sigma}$ increases. In
this case, the sublattice magnetic moments have more prominent
variations with the change of the couplings than the previous cases.

\begin{figure}[tbp]
\includegraphics[width=1.0\linewidth,clip]{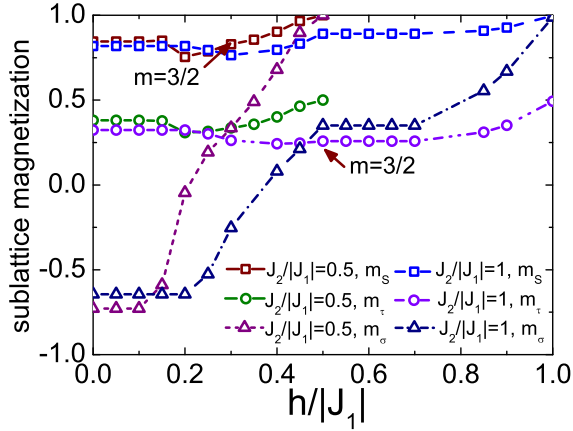}
\caption{(Color online) Magnetic field dependence of the sublattice
magnetization $m_{\mathrm{S}}$, $m_{\mathrm{\tau}}$ and
$m_{\mathrm{\sigma}}$ for $J_{2}$/$\vert J_{1}\vert$=$0.5$ and $1$.
The arrows indicate the magnetic field $h_{c2}$ where the
magnetization per unit $m$=$3/2$.} \label{Mag2}
\end{figure}

In the $m$=$\frac{3}{2}$ plateau ($h_{c2}$$<$$h$$<$$h_{c3}$), the
local magnetic moments for $J_{2}$/$\vert J_{1}\vert$=1 are shown in
Fig. \ref{F-AF}(d). By comparing the local magnetization below
$h_{c1}$ [Fig. \ref{F-AF}(c)] and in the plateau [Fig.
\ref{F-AF}(d)], it is found that the novel decreasing feature of
magnetization $m_{\tau}$ found in case (B) is also observed in the
present case. The field dependence of sublattice magnetization is
also studied. As illustrated in Fig. \ref{Mag2} for $J_{2}$/$\vert
J_{1}\vert$=$0.5$ and $1.0$, both $m_{\tau}$ and $m_{\mathrm{S}}$
have decreasing regions from $h_{c1}$ to $h_{c2}$. Comparing with
the sublattice magnetization of case (B), we notice that the two
sublattice magnetizations that have a decreasing region from
$h_{c1}$ and $h_{c2}$ are those coupled by FM interactions, and the
decreasing behavior is only observed below the $m$=$\frac{3}{2}$
plateau.

\section{Temperature dependence of susceptibility and specific heat}

From the above results, it can be seen that although the three cases
entirely exhibit FI ground states, the low-lying excitations and
magnetic properties are rather distinct. Thus, in this section, the
temperature dependences of zero-field magnetic susceptibility and
specific heat are explored by the TMRG method.\cite{TMRG} In the
following calculations, the width of the imaginary time slice is
taken as $\varepsilon$=$0.1$, and the error caused by the
Trotter-Suzuki decomposition is less than $10^{-3}$. During the TMRG
iterations, $120$ and $200$ states are retained for the evaluation
of the susceptibility and specific heat, respectively, and the
temperature is down to $k_{B}T$=$0.025\vert J_{1}\vert$ in general.
The truncation error is less than $10^{-4}$ in all calculations.

\begin{figure}[tbp]
\includegraphics[width=1.0\linewidth,clip]{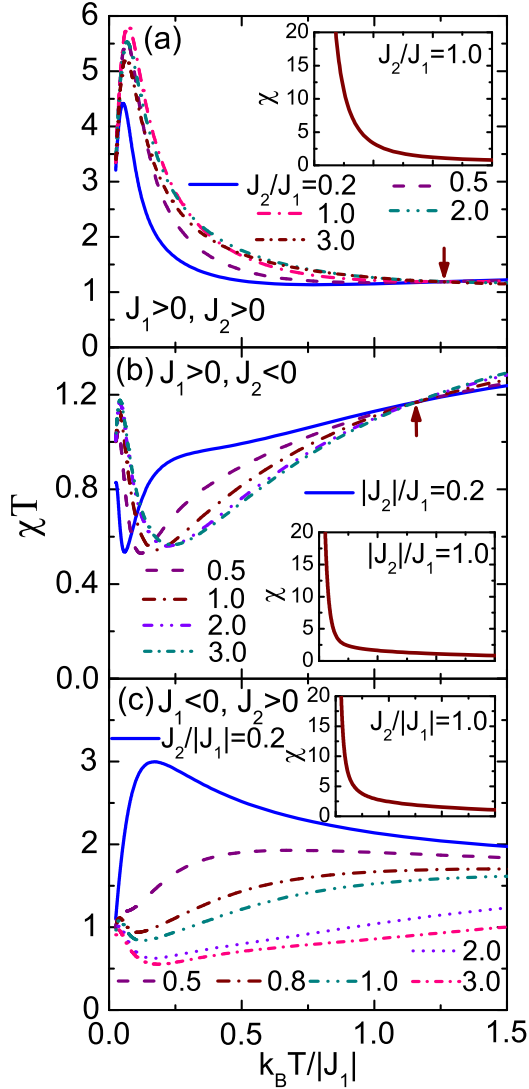}
\caption{(Color online) Temperature dependence of $\chi T$ for (a)
$J_{1}$$>$$0$, $J_{2}$$>$$0$; (b) $J_{1}$$>$$0$, $J_{2}$$<$$0$; and
(c) $J_{1}$$<$$0$, $J_{2}$$>$$0$. The insets show the susceptibility
as a function of temperature.}
\label{Thermal}
\end{figure}

The temperature dependence of the susceptibility $\chi$ and
susceptibility temperature product $\chi T$ for the cases are shown
in Figs. \ref{Thermal}(a)-(c). For case (A), the susceptibility, as
shown in the inset of Fig. \ref{Thermal}(a), diverges as
$T$$\rightarrow$$0$ due to the gapless branch $\omega_{1,k}$ [Fig.
\ref{AF-AF}(a)]. Upon lowering temperature, $\chi T$ decreases to a
broad minumum at a temperature $T_{\mathrm{min}}$, and then
increases to a peak at low temperauture $T_{\mathrm{peak}}$. The
minimum of $\chi T$ is an indicative of the FI-like behavior similar
to that in the spin-($\frac{1}{2},1$) mixed-spin chain. With
increasing $J_{2}/J_{1}$, $T_{\mathrm{min}}$ shifts to higher
temperatures, corresponding to the enhancement of the branches
$\omega_{2,k}$ and $\omega_{3,k}$ with the increase of coupling
ratios. Meanwhile, $\chi T_{\mathrm{min}}$ increases for
$J_{2}/J_{1}$$<$$1$ and decreases for $J_{2}/J_{1}$$>$$1$. The
maximum of $\chi T_{\mathrm{min}}$ is reached at
$T_{\mathrm{min}}$=$1.25J_{1}$ when $J_{2}/J_{1}$=$1$. It is also
noticed that the $\chi T$ curves for different couplings intersect
at the same temperature $1.25J_{1}$, as shown by the arrow in Fig.
\ref{Thermal}(a). At low temperature, $\chi T$ does not diverge,
like that in the spin-($\frac{1}{2},1$) mixed-spin chain\cite{QMC},
but has a sharp peak, which indicates that $\chi$ diverges equally
or slower than $\frac{1}{T}$ as $T$$\rightarrow$$0$.\cite{Chi} For
the low-temperature peak, it is unveiled that $T_{\mathrm{peak}}$
moves to higher temperatures with the increase of the height for
$J_{2}/J_{1}$$<$$1$, while it approaches lower temperatures with the
height decreasing for $J_{2}/J_{1}$$>$$1$. It can be seen that the
finite-temperature magnetic properties have transition behaviors
with the change of the couplings at $J_{2}/J_{1}$=$1$, which was
also noted in the ground states in Sec.
\uppercase\expandafter{\romannumeral4}(A).

For case (B), $\chi$ also goes to infinity as $T$$\rightarrow$$0$
owing to the gapless branch $\omega_{1,k}$ [the inset of Fig.
\ref{Thermal}(b)]. As shown in Fig. \ref{Thermal}(b), $\chi T$
decreases rapidly to a minimum at $T_{\mathrm{min}}$ with decreasing
temperature, and then increases to a peak at lower temperature,
which is quite different from that in the case (A). With increasing
$\vert J_{2}\vert/J_{1}$, both $T_{\mathrm{min}}$ and $\chi
T_{\mathrm{min}}$ enhance. As indicated by the arrow in Fig.
\ref{Thermal}(b), $\chi T$ curves for different couplings also
intersect at a temperature $T$$\sim$$1.15J_{1}$. At low temperature,
both the peak temperature and height of $\chi T$ increase with
increasing $\vert J_{2}\vert/J_{1}$. The peak suggests that $\chi$
diverges equally or slower than $\frac{1}{T}$ as
$T$$\rightarrow$$0$. Different from the case (A), the variation of
$\chi T$ in the present case with FM coupled pendants does not
exhibit a transition behavior. It can be seen that $J_{2}$ has a
great impact on the low-lying excitations as well as the magnetic
properties at finite temperature.

Figure \ref{Thermal}(c) illustrates the behaviors of $\chi T$ for
case (C). Although $\chi$ also diverges as $T$$\rightarrow$$0$ [the
inset of Fig. \ref{Thermal}(c)], $\chi T$ has rather distinct
behaviors from cases (B) and (C) with $J_{1}$$>$$0$. For
$J_{2}/\vert J_{1}\vert$=$0.2$, $\chi T$ increases to a broad
maximum with decreasing temperature, and then declines. When
$J_{2}/\vert J_{1}\vert$$>$$0.5$, a minimum of $\chi T$ emerges, and
a small peak appears at a lower temperature. For $J_{2}/\vert
J_{1}\vert$$>$$1$, $\chi T$ decreases to a minimum with declining
temperature, showing the AFM feature, and then increases to a small
peak, which is similar to that in the case (B). The minimum
temperature $T_{\mathrm{min}}$ also increases with enhancing
$J_{2}/\vert J_{1}\vert$. The convergence of $\chi T$ as
$T$$\rightarrow$$0$ indicates that $\chi$ diverges equally or slower
than $\frac{1}{T}$ as $T$$\rightarrow$$0$ in this case. Compared
with the above cases, no intersection of $\chi T$ is observed for
the present case with $J_{1}$$<$$0$. It should be noted that the
similar behavior of $\chi T$ has also been observed in the
spin-($\frac{1}{2},1$) AFM chain with FM next-nearest-neighbor
coupling,\cite{SSKS} which has an analogous low-lying excitations.
As shown in Fig. \ref{F-AF}(a), the gapped magnon branch
$\omega_{2,k}$ has lower energies than the gapless branch
$\omega_{1,k}$ for large wave momenta $k$. Thus, the low-lying
excitations are dominated by $\omega_{2,k}$ for small $J_{2}$. With
increasing $J_{2}/\vert J_{1}\vert$, $\omega_{2,k}$ enhances and
$\omega_{1,k}$ gradually dominates the low-lying excitations. The
branches $\omega_{1,k}$ and $\omega_{2,k}$ become analogous to that
of the case (B) [Fig. \ref{AF-F}(a)] for large $J_{2}/\vert
J_{1}\vert$, yielding the behaviors of $\chi T$ for $J_{2}/\vert
J_{1}\vert$$>$$1$ similar to that of the case (B) [Fig.
\ref{Thermal}(b)].

\begin{figure}[tbp]
\includegraphics[width=1.0\linewidth,clip]{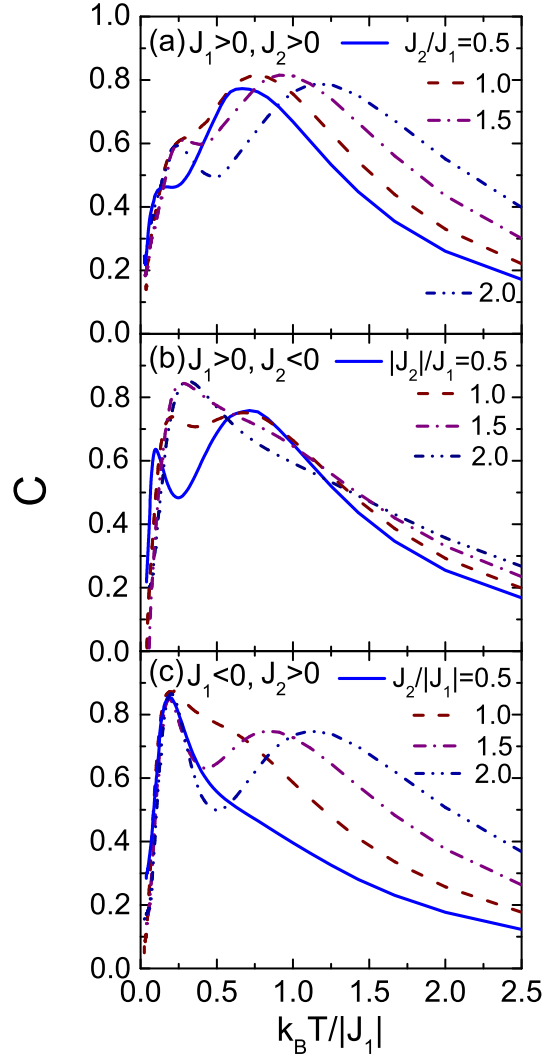}
\caption{(Color online) Temperature dependence of the specific heat $C$ for (a)
$J_{1}$$>$$0$, $J_{2}$$>$$0$; (b) $J_{1}$$>$$0$, $J_{2}$$<$$0$; and
(c) $J_{1}$$<$$0$, $J_{2}$$>$$0$.}
\label{ThermalCv}
\end{figure}

In Figs. \ref{ThermalCv}(a)-(c), the temperature dependences of the
specific heat for the three cases are shown explicitly. For case
(A), the specific heat has a prominent double-peak structure. When
$J_{2}/J_{1}$=$0.5$, the high-temperature peak of specific heat is
close to the peak temperature of that in the spin-($\frac{1}{2},1$)
mixed-spin chain. With further increasing $J_{2}$, the
low-temperature peak shifts to higher temperatures when
$J_{2}$/$J_{1}$$<$$1$, while it keeps nearly intact for
$J_{2}$/$J_{1}$$>$$1$. Meanwhile, the high-temperature peak
continuously moves to higher temperatures, which might be owing to
the enhancement of the gapped branch $\omega_{2,k}$ and
$\omega_{3,k}$.

The temperature dependence of specific heat for case (B) is shown in
Fig. \ref{ThermalCv}(b). When $\vert J_{2}\vert/J_{1}$=$0.5$, the
specific heat has double peaks, and the high-temperature peak is
also close to the peak temperature of that in the
spin-($\frac{1}{2},1$) mixed-spin chain. Compared with the specific
heat of the case (A) with $J_{2}$/$J_{1}$=$0.5$ [Fig.
\ref{ThermalCv}(a)], it can be seen that the high temperature
behaviors above the high-temperature peak of the two cases agree
well with each other, but the low-temperature behaviors are
distinct. With increasing $\vert J_{2}\vert/J_{1}$ for $\vert
J_{2}\vert/J_{1}$$<$$1$, the low-temperature peak moves to higher
temperature side, while the high-temperature peak keeps nearly
intact. For $\vert J_{2}\vert/J_{1}$$>$$1$, the double peaks merge
into a single peak, which moves to higher temperatures slightly with
increasing $\vert J_{2}\vert/J_{1}$. Analogous to $\chi T$, $J_{2}$
has also an essential effect on the behaviors of specific heat.

For case (C), the specific heat behaves quite differently from the
above cases. For $J_{2}$/$\vert J_{1}\vert$=$0.5$, the specific heat
shows a single peak instead of double peaks at low temperature. With
increasing $J_{2}$ below $J_{2}$/$\vert J_{1}\vert$=$1$, the
specific heat below the peak temperature keeps nearly unchanged,
while the part above the peak temperature decreases more slowly, as
shown in Fig. \ref{ThermalCv}(c). For $J_{2}$/$\vert
J_{1}\vert$$>$$1$, a high-temperature peak emerges, which moves to
higher temperatures with increasing $J_{2}$/$\vert J_{1}\vert$.
Meanwhile, the behaviors of specific heat below the low-temperature
peak still retains nearly intact. The low-temperature peak seems to
be insensitive to $J_{2}$, and is dominated by $J_{1}$.

For a comparison, we also calculated the thermal quantities by the
LSW theory, which give rise to the similar behaviors for the three
different cases. The results show that $\chi T$ diverges as
$T$$\rightarrow$$0$, and the specific heat always exhibits double
peaks. Although the obtained low-lying excitations are helpful to
understand the thermodynamics, the quantitative results obtained
from the LSW are not so good, which are thus not presented here.

\section{Summary and Discussion}

In this paper, the low-lying, magnetic and thermodynamic properties
of the spin-($\frac{1}{2},1$) decorated mixed-spin chain with
spin-$1$ pendant spins are systematically studied for three cases:
(A) $J_{1},J_{2}$$>$$0$; (B) $J_{1}$$>$$0$, $J_{2}$$<$$0$; and (C)
$J_{1}$$<$$0$, $J_{2}$$>$$0$ by jointly using a few different
methods.

By means of the RSRG analysis, the low-energy effective Hamiltonians
for each cases in strong and weak couplings are obtained. It is
found that although the effective Hamiltonians are different for
three cases, their magnon excitations from $S_{G}$ to $S_{G}-1$ are
all FM and gapless, which agree with that of the
spin-($\frac{1}{2},1$) mixed-spin chain without pendants. The
low-energy dispersions of the gapless branch near $k$=$0$ are
dominated by $J_{1}$ for each case, which is confirmed by the LSW
results.

The low-lying excitations and magnetic properties are then
investigated by the LSW and DMRG methods, respectively. The magnon
spectra are found to consist of a gapless and a gapped branches from
$S_{G}$ to $S_{G}-1$, as well as a gapped branch from $S_{G}$ to
$S_{G}+1$, which have different features for three cases. For case
(C), two low-energy branches have a novel intersection. In a
magnetic field, case (A) has a $m$=$\frac{3}{2}$ plateau, while both
cases (B) and (C) exhibit two plateaux at $m$=$\frac{1}{2}$ and
$\frac{3}{2}$. The low-energy gap of case (A) increases almost
linearly with increasing the coupling ratio, while those of cases
(B) and (C) increase and go to saturation for large $\vert
J_{2}\vert$, which implies that the low-energy gap of the case (A)
is mainly scaled by $J_{2}$, and those of the cases (B) and (C) are
scaled by $J_{1}$ for large $\vert J_{2}\vert$. The sublattice
magnetization of the spins coupled by FM interactions for cases (B)
and (C) are found to decrease in some regions from $h_{c1}$ to
$h_{c2}$ with the increase of the magnetic field, which may be
attributed to the competition of the AFM and FM interactions in a
magnetic field.

The zero-field thermodynamics are also explored by means of the TMRG
method. It is unveiled that although $\chi$ diverges as
$T$$\rightarrow$$0$, $\chi T$ has rather different behaviors for
each cases. For case (A), $\chi T$ has a broad minimum and a peak at
low temperature. The curves of $\chi T$ for different couplings
intersect at a common temperature $1.25J_{1}$, and $\chi T$ has a
transition behavior with the couplings at $J_{2}$/$J_{1}$=$1$. For
case (B), $\chi T$ has a narrow minimum and a sharp peak at low
temperature. The curves of $\chi T$ for different couplings also
intersect at a common temperature, but $\chi T$ never show a
crossing behavior. For case (C), $\chi T$ has a broad peak for
$J_{2}$/$\vert J_{1}\vert$$<$$1$, and exhibits a broad minimum and a
peak for $J_{2}$/$\vert J_{1}\vert$$>$$1$, showing two distinct
features with changing the couplings due to the intersection of two
low-lying excitations. Compared with the spin-($\frac{1}{2},1$)
mixed-spin chain, there is a common feature for the three cases that
$\chi T$ converges as $T$$\rightarrow$$0$, which implies that $\chi$
diverges equally or slower than $\frac{1}{T}$ as
$T$$\rightarrow$$0$.

The specific heat for case (A) has double peaks. For case (B), the
specific heat has double peaks when $\vert
J_{2}\vert$/$J_{1}$$<$$1$, which merge into a single peak as $\vert
J_{2}\vert$/$J_{1}$$>$$1$. For case (C), however, the specific heat
has a single peak when $J_{2}$/$\vert J_{1}\vert$$<$$1$, while
double peaks emerge when $J_{2}$/$\vert J_{1}\vert$$>$$1$. In a wide
range of the coupling for case (C), the low-temperature peak appears
to be insensitive to $J_{2}$, which mainly affects the
high-temperature behaviors of the specific heat.

Based on the above results, it can be seen that the case (A) of the
present system preserves some features of the spin-($\frac{1}{2},1$)
mixed-spin chain, while the cases (B) and (C) exhibit more novel
exotic properties that have not been observed in the mixed-spin
chains. We expect that the magnetic and thermodynamic properties
presented in this paper could be tested experimentally in future to
unveil the effects induced by the pendant spins in the mixed-spin
chains.

\acknowledgments
We are grateful to Hui-Zhong Kou for helpful
discussions. The work is supported in part by the NSFC (Grant Nos.
10625419, 10934008, 90922033), the MOST of China (Grant No.
2006CB601102) and the Chinese Academy of Sciences.


\begin{thebibliography}{99}
\bibitem{Chi} In order to investigate the low-temperature behavior of $\chi T$,
we have performed the TMRG calculations to lower temperatures.
However, as $\chi T$ decreases rather sharply as
$T$$\rightarrow$$0$, it is hard to determine exactly whether $\chi
T$ converges to zero or just to a finite small value.
\end{thebibliography}
\end{document}